\documentclass[aps,prl,twocolumn,altaffilletter,longbibliography,numerical,flushbottom,secnumarabic,superscriptaddress,floatfix,nobibnotes]{revtex4-2}

\usepackage{graphicx}
\usepackage{braket}
\usepackage{amsmath, amssymb}
\usepackage{booktabs}  
\usepackage{comment}
\usepackage[normalem]{ulem}
\usepackage{cancel}

\usepackage[breaklinks, pdftex, hyperfootnotes=true, pdfpagelabels, bookmarks, pageanchor]{hyperref}
\hypersetup{%
	colorlinks=true, linktocpage=true, pdfstartpage=1, pdfstartview=FitH, pdfborder={0 0 0},%
	breaklinks=true, pdfpagemode=UseNone, pageanchor=true, pdfpagemode=UseOutlines,%
	plainpages=false, bookmarksnumbered, bookmarksopen=true, bookmarksopenlevel=1,%
	hypertexnames=true, pdfhighlight=/O,
	urlcolor=blue, linkcolor=blue, citecolor=blue,
	}
\usepackage{orcidlink}

\newcommand\identity{1\kern-0.25em\text{l}}

\graphicspath{{figs/}}
 
\usepackage{soul}

\usepackage[colorinlistoftodos]{todonotes}
\presetkeys{todonotes}{inline}{}

\begin{document}

\title{Antichiral edge states and Bogoliubov Fermi surfaces in a two-dimensional proximity-induced superconductor}

\newcommand{\affA}{\affiliation{Departamento de S\'olidos Cu\'anticos y Sistemas Desordenados, Centro At\'omico Bariloche, Instituto de Nanociencia y Nanotecnolog\'ia CONICET-CNEA and Instituto Balseiro (8400), San Carlos de Bariloche, Argentina.}}
\newcommand{\affB}{\affiliation{Grupo de Circuitos Cuánticos Bariloche, Div. Dispositivos y Sensores, Centro Atómico Bariloche-CNEA, Instituto Balseiro and CONICET, (8400) San Carlos de Bariloche, Argentina.}}
\newcommand{\affC}{\affiliation{Instituto de Ciencias Físicas y Escuela de Ciencia y Tecnología, Universidad Nacional de San Martín, (1650) Buenos Aires, Argentina.}}
\newcommand{\affD}{\affiliation{Institut f\"ur Experimentelle und Angewandte Physik, University of Regensburg, 93040 Regensburg, Germany.}}

\author{Gabriel F. Rodríguez Ruiz~\orcidlink{0000-0002-0425-3772}}
\affA
\author{Juan Herrera Mateos~\orcidlink{0000-0001-7107-2077}}
\affC
\author{Leandro Tosi}
\affB
\author{Christoph Strunk}
\affD
\author{Carlos Balseiro}
\affA
\author{Liliana Arrachea~\orcidlink{0000-0002-7223-4610}}
\email[Corresponding author: ]{liliana.arrachea@ib.edu.ar}
\affA

\date{\today} 

\begin{abstract}
 We show that a magnetic field parallel to the plane of a two-dimensional electron gas with Rashba spin orbit coupling in proximity to a superconductor leads to a topological phase
 in coexistence with a single pair of Bogoliubov Fermi surfaces. This phase hosts antichiral edge states of co-propagating Majorana fermions and are spatially localized at the opposite edges of the sample, perpendicular to the magnetic field.  We discuss the characteristic signatures in the current-phase relation of a Josephson junction formed by two reservoirs in the topological phase.
\end{abstract}

\maketitle

{\em Introduction. } The search for Majorana zero modes in solid-state devices has driven intense efforts to realize topological superconductivity. Most approaches engineer hybrid platforms combining conventional s-wave superconductors, spin-orbit coupling (SOC) and time-reversal-symmetry breaking mechanism, like a magnetic field, magnetic properties of materials \cite{Alicea2012Jun,Prada2020Oct,Flensberg2021Oct,Schiela2024Sep}. Heterostructures of semiconducting InAs and InSb, having a large g-factor and sizable spin-orbit coupling offer a natural playground in this context.
Wires fabricated in these materials captured most of the theoretical \cite{Lutchyn2010Aug,Oreg2010Oct} and experimental early attention. This topological phase has been also explored in engineered two-dimensional (2D) configurations \cite{Kjaergaard2016Sep}, and prominent examples are planar Josephson junctions  \cite{Pientka2017May,Hart2017Jan,Fornieri2019May,Ren2019May,Banerjee2023Jun}.  Beyond the realization of the localized Majorana modes, 2D superconductors with SOC and broken time-reversal symmetry also exhibit rich physics  \cite{Moehle2021Dec},
like the superconducting diode effect \cite{Ando2020Aug,BaumgartnerPRL2021,BaumgartnerNature2022,Banerjee2023Nov}, the formation of Bogoliubov Fermi surfaces \cite{Agterberg2017Mar,Yuan2018Mar,setty2020bogoliubov,Shaffer2020Jun,Dutta2021Sep,Phan2022Mar,Babkin2024May,Mateos2024Aug,Pal2024Dec,Cohen2024Apr,Ohashi2024Sep}  and the realization of the Fulde–Ferrell–Larkin–Ovchinnikov state \cite{Mandal2024Jun} are equally fascinating.  

In this work we show the existence of a new topological superconducting phase in a 2D electron system with SOC when a magnetic field is applied in-plane. The bulk spectrum of this phase is not fully gapped but has a structure of two Dirac-like cones and hosts antichiral Majorana modes that co-propagate along the edges perpendicular to the magnetic field. This feature is in contrast to fully gapped 2D topological superconductors with broken time-reversal symmetry generated by applying a magnetic field perpendicular to the plane, which has
Majorana modes propagating clockwise or anticlockwise along the edges \cite{Sau2010Jan,Alicea2010Mar,Bernevig2013Apr,Sato2017May}. Such phase is characterized by a Chern number and  has been found to be robust against a tilt of the magnetic field with a finite in-plane projection \cite{Loder2015Oct}. 

\begin{figure}[t!]
\includegraphics[width=.49\textwidth]{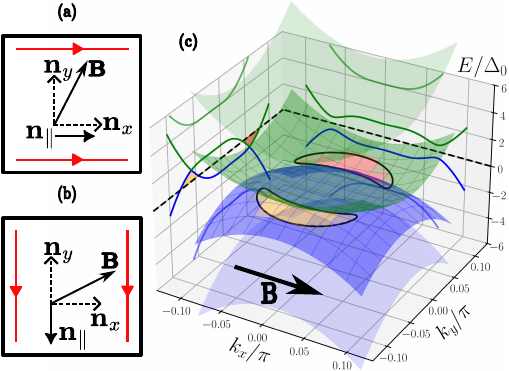} 
\caption{\textbf{Antichiral edge states:} A 2D electron system in proximity with an $s$-wave superconductor is placed in the $x,y$ plane. The Rashba SOC is in the plane as well as the external magnetic field ${\bf B}=(B_x,B_y,0)$. A topological phase with antichiral edge states along ${\bf n}_{||}$  exists for $\mu^2 \leq V^2-\Delta_0^2$ within a range of angles satisfying $|{\bf n}_{||} \cdot  {\bf n}_V|< \Delta_0/V<1$ (see text). \textbf{(a)} and \textbf{(b)} illustrate two different configurations of the magnetic field and the antichiral edge states. The existence of the topological phase is accompanied by the emergence of a single pair of Bogoliubov Fermi Surfaces in the spectrum.  \textbf{(c)} BdG spectrum obtained within the lattice model for the magnetic field oriented along ${\bf n}_x$. The projected plane-cuts correspond to $(k_x=0,k_y,E)$ and $(k_x,k_y=0,E)$ surfaces. 
}
\label{fig:spec}
\end{figure}

Topological systems with antichiral states were theoretically predicted in a modified Haldane model \cite{Colomes2018Feb} and experimental realizations have been reported in photonic systems \cite{Zhou2020Dec,Xi2023Apr,Wang2021Feb,Chen2022Jun,Hu2024Nov}. Similar  co-propagating states were predicted to emerge due to the coupling between electrons and chiral phonons in graphene lattices \cite{MedinaDuenas2022Feb,Mella2023Dec} and models for altermagnets\cite{Sorn2025Apr}. In contrast to these examples, 
the edge modes of the 2D superconducting phase we discuss in the present work are akin to lines of Majorana zero modes of stacked topological wires. This type of topological phase, where the boundary modes are present only in specific edges is identified as ``weak topological phases'' and have been mainly studied in the context of insulators \cite{Ringel2012Jul} and semimetals \cite{Wieder2020Jan}. As mentioned in these studies \cite{Ringel2012Jul}, the name does not mean lack of robustness for the edge states. We discuss the spectral properties and the topological characterization of this phase. Very interestingly, it coexists with the development of a single pair of Bogoliubov Fermi surfaces. We focus on parameters similar to those of Al/InAs heterostructures \cite{BaumgartnerPRL2021,BaumgartnerNature2022} and discuss  the experimental signatures of this phase in the current-phase relation (CPR) of Josephson junctions tailored in such 2D platform. 

{\em Model.} A sketch of the system is presented in Fig. \ref{fig:spec}. We express the Hamiltonian for the 2D electron system with  SOC and magnetic field in the basis ${\bf c}_{\bf k}=\left(c_{\bf k,\uparrow},c_{{\bf k},\downarrow}\right)^T$. It reads ${\cal H}_{\rm 2D}({\bf k})={\bf c}_{\bf k}^\dagger\left[  \xi_{\bf k}  +  H_{\rm SOC}({\bf k})+H_{\rm Z}\right]\;{\bf c}_{\bf k}$, where the first term is the kinetic dispersion relation and the next terms are the SOC and Zeeman Hamiltonians,
\begin{equation}\label{hcont1}
    H_{\rm SOC}({\bf k})=-  {\bf n}_z \cdot \left({\boldsymbol \sigma} \times {\boldsymbol \lambda}_{\bf k}\right),\;\;\;\;\;\;\;\;\; H_{\rm Z}=-V {\bf n}_V \cdot {\boldsymbol \sigma}.
\end{equation}
The Zeeman field is $V=\frac{1}{2}g \mu_B B $, being $g$ the g-factor and $B$ the strength of the magnetic field applied in the direction ${\bf n}_V$, while ${\boldsymbol \sigma}$ is the vector of Pauli matrices. In a continuum model the kinetic dispersion relation is $\xi_{\bf k}=k^2/2m -\mu$, being $\mu$ the chemical potential. We focus on Rashba SOC with  ${\boldsymbol \lambda}_{\bf k}=\alpha_R \left(k_x,k_y, 0 \right)$. In the calculations we also consider a square-lattice model for this system with hopping parameter $t$, in which case $\xi_{\bf k}= -2 t \left(\cos k_x + \cos k_y\right)-\mu+4t$ and the SOC is ${\boldsymbol \lambda}_{\bf k}=2 \lambda\left(\sin k_x, \sin k_y, 0 \right)$.  The proximity to the superconducting layer is modeled by a local s-wave pairing with strength $\Delta_0$ described by the Hamiltonian
\begin{equation}\label{pairing0}
{\cal H}_{\Delta}({\bf k})=  -\frac{\Delta_{0}}{2} \left(c^{\dagger}_{{\bf k},\uparrow}c^{\dagger}_{-{\bf k},\downarrow}
+ c^{\dagger}_{-{\bf k},\uparrow}c^{\dagger}_{{\bf k},\downarrow}+H. c.\right).
\end{equation}
The ensuing Bogoliubov-de-Gennes (BdG) Hamiltonian expressed in the Nambu basis $\Psi_{\bf k}=\left(c_{\bf k,\uparrow},c_{{\bf k},\downarrow},c^\dagger_{-{\bf k},\downarrow},-c^\dagger_{-{\bf k},\uparrow}\right)^T$ reads
\begin{equation}\label{bdg-cont}
    H_{\rm BdG}({\bf k})= \frac{1}{2}\left[\xi_{\bf k}+H_{\rm SOC}({\bf k})\right] \; \tau^z + \frac{1}{2} H_{\rm Z} -\frac{\Delta_0}{2} \; \tau^x,
\end{equation}
where the Pauli matrices $\tau^j$ act on the particle-hole degree of freedom. 

A crucial ingredient for the topological phase we study is a magnetic field with a non-vanishing projection along the direction of the SOC. We hereafter focus on a fully in-plane magnetic field, as indicated in the schemes of Fig. \ref{fig:spec}(a, b). In what follows we consider ${\bf n}_V\equiv {\bf n}_x$, which implies breaking the rotational symmetry in the plane. Hence, the spectrum has completely different features along the directions ${\bf n}_x$ (parallel) and ${\bf n}_y$ (perpendicular) to the magnetic field, respectively. It is important to  notice that for $k_y=0$, the spin orientation of the SOC is along ${\bf n}_y$ and the Hamiltonian Eq. (\ref{bdg-cont}) reduces to the model of helical wires originally proposed in Refs. \cite{Lutchyn2010Aug,Oreg2010Oct}. This model has a topological phase for $\mu^2 \leq V^2-\Delta_0^2$. Fig. \ref{fig:spec}(c) illustrates the bulk spectrum of the 2D model within this range of parameters. It can be observed that while the spectrum along $k_x$ is gapped, along $k_y$ two cones with a quadratic dispersion relation develop close to $k_y=0$. They cross zero energy for larger values, leading to the formation of a single pair of Bogoliubov Fermi surfaces, as already pointed out in Refs. \cite{Yuan2018Mar,Babkin2024May}. These cones are asymmetric in the sense that one of them intersects the zero-energy plane from below, while the other from above. 

\begin{figure}[t]
\includegraphics[width=.5\textwidth, trim = 0.3cm 0.6cm 0 0.3cm, clip]{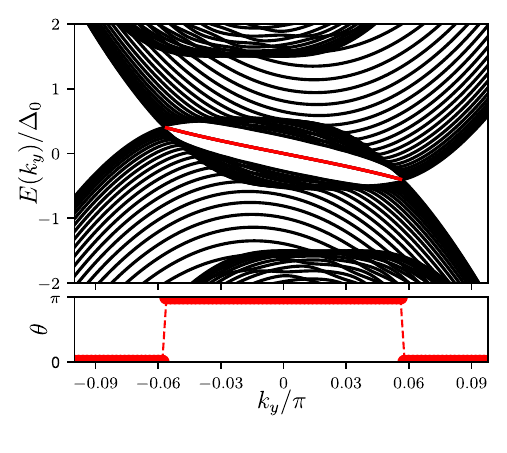} 
\caption{\textbf{Edge states and topological invariant:} \textbf{(a)} Bogoliubov-de Gennes spectrum of a ribbon of $N_x=200$ lattice sites  with open boundary conditions (OBC) in $x$ and
periodic boundary conditions (PBC) along $y$, as a function of $k_y$ in the topological phase for the configuration of Fig. \ref{fig:spec}(c), magnetic field along ${\bf n}_x$. The spectrum shows the dispersion relation of the edge states connecting the cones above and below the Fermi level. \textbf{(b)} Topological invariant $\theta (k_y)$ obtained from the numerical evaluation of the eigenvalues of the Wilson loop defined in Eq. (\ref{eq:wilson_loop})(see text). Parameters are: $t=50\Delta_0$, $\lambda=2.8\Delta_0$, $V=(2\Delta_0,0,0)$, $\mu=0$.}
\label{fig:Wilson-loop}
\end{figure}

{\em Topological properties.} The system we consider belongs to the class D of the classification presented in Ref. \cite{Schnyder2009May}. In fact, it has time-reversal symmetry broken while preserves particle-hole symmetry. In the case of the 1D system corresponding to $k_y=0$, the topological properties are defined by the value of a Berry-Zak phase and by the existence of Majorana zero modes localized at the ends of the wire. In the 2D system we study, the appropriate invariant is 
similar to that introduced to characterize Dirac semimetals and high-order topology  \cite{Fidkowski2011Jul,Alexandradinata2014Apr,Cano2018Jun,Bradlyn2019Jan,Bouhon2019Nov,Wieder2020Jan,Wang2019Sep,Awoga2022Apr}. We consider periodic boundary conditions along $x$ and $y$ and calculate the $x$-directed Wilson loop (holonomy) matrix  with elements
\begin{equation}
    {\cal W}^{\ell,\ell'}(k_y)=P\exp\left\{i \int_0^{2\pi}dk_x A^{\ell,\ell'}_{k_x}(k_y)\right\},
    \label{eq:wilson_loop}
\end{equation}
where $P$ denotes path ordering taken over the Brillouin zone of the one-dimensional Hamiltonian with fixed $k_y$. 
This generalizes the Zak phase along $x$ for a fixed value of $k_y$.  In the present case, the
Berry connection is non-abelian and is defined as $A^{\ell,\ell'}_{k_x}(k_y) =i \langle u^{\ell}_{\bf k}|\partial_{k_x}u^{\ell'}_{\bf k} \rangle$. The indices 
$\ell$, $\ell'$ label the
two lowest-energy eigenstates  of the Hamiltonian in Eq. (\ref{bdg-cont}). We evaluate ${\cal W}$ by   discretizing the path as in Refs. \cite{Ortiz1996Nov,PerezDaroca2021Sep}, 
 then  diagonalize it  to obtain the eigenvalues $\theta_\ell(k_y)$.
 The phase $\theta(k_y)=\sum_{\ell} \theta_\ell(k_y)=\pi, 0$ defines the topological characterization\cite{Alexandradinata2014Apr,Cano2018Jun,Bradlyn2019Jan,Fidkowski2011Jul,Wieder2020Jan}.
  This, respectively, indicates the existence or 
non existence of edge
states for that $k_y$.
In Fig. \ref{fig:Wilson-loop}(b) we show the result for $\theta(k_y)$ calculated with the parameters which give the spectrum of panel (a). The energies as a function of $k_y$ are obtained for a ribbon with $N_x=200$ lattice sites with open boundary conditions (OBC) along $x$
and periodic boundary conditions (PBC) along $y$. We can clearly identify the two cones of the bulk states, associated to the formation of a single pair of Bogoliubov Fermi surfaces, confining the range of $k_y$ within which
$\theta(k_y)=\pi$. In addition to the bulk states, we identify doubly degenerate edge states depicted in red. They are effectively described by the Hamiltonian
\begin{equation}\label{h-edge}
    {\cal H}_{\rm edge} = \sum_{\nu=l,r; k_y \geq 0} v  k_y \; \eta^{\dagger}_{\nu,k_y} \eta_{\nu,k_y},
\end{equation}
where $\eta^{\dagger}_{\nu,k_y}=\eta_{\nu,-k_y}$ are Majorana modes that propagate along the $y$ direction with the same velocity $v$, while they are spatially localized at the left and right ($\nu=l,r$) edges of the ribbon. Therefore, these modes are {\rm antichiral} and include the Majorana zero modes $\eta^{\dagger}_{\nu,0}=\eta_{\nu,0}$ of the topological 1D limit.  
The velocity is defined by the component of the SOC perpendicular to the magnetic field and we
present below an approximate expression on the basis of a low-energy Hamiltonian. Akin to the models of  photonic systems \cite{Colomes2018Feb,Zhou2020Dec,Chen2022Jun}, these modes transport energy in one way along the edges, while there is a
counterflow through the bulk states.

\begin{figure}[t!]
\includegraphics[width=.5\textwidth, trim={0 1.55cm 0cm 0cm},clip]{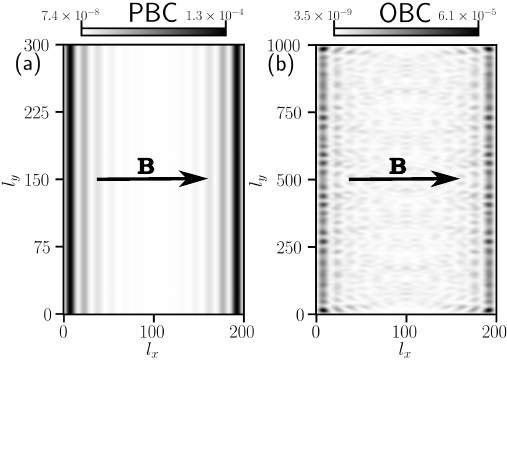}
\caption{\textbf{Mode localization and robustness:} Spatial probability distribution of the lowest positive energy eigenstate $E\approx 0$ within the topological phase,  for a lattice with $N_x\times N_y$ sites (labeled with $l_x,l_y$). \textbf{(a)} OBC  in the $x$-direction and PBC in the $y$-direction. \textbf{(b)} OBC are considered in both directions. In both cases $N_x=200$ sites. In (a), $N_y^{(PBC)}=300$, $E/t=9.1\times10^{-5}$. In (b), $N_y^{(OBC)}=1000$ and $E/t=5.6\times10^{-5}$. Other 
parameters are the same as in Fig. \ref{fig:Wilson-loop}.
}
\label{fig:prob-edge}
\end{figure}

The spatial probability distribution of the lowest-positive-energy antichiral edge states is shown in Fig. \ref{fig:prob-edge}. In both panels, the ribbon is described with OBC in $x$. Panel (a) shows results obtained with PBC in $y$, where the two edge states confined at the $l$ and $r$ sides of the ribbon can be clearly distinguished. Because this analysis assumes  translational invariance along the $y$ direction, the question now arises on the robustness of these modes when this symmetry is broken. In panel (b) the spatial probability density is calculated with open boundary conditions in the two directions. The effect of the OBC is to introduce interference effects in the edge states mediated by the bulk states but it does not prevent their formation. Similar features are observed in a system with circular shape (see Ref. \cite{sm}).

So far, we have considered a fixed orientation of the magnetic field, perpendicular to the ribbon. We would like to stress that the topological phase is not limited to this particular configuration. We recall that the $k_y=0$ channel is equivalent to the 1D system and hence, we can rely on the  boundaries of the topological phase provided in this limit by  Refs. \cite{Osca2014Jun,Rex2014Sep,Klinovaja2015Mar, Aligia2020Dec,Gruneiro2023Jul} for arbitrary orientations of the magnetic field. The result is 
 \begin{equation} \label{angle}
  |{\bf n}_{||} \cdot  {\bf n}_V|< \Delta_0/V<1,
 \end{equation}
 where ${\bf n}_{||}$ is the orientation of the edge that hosts the antichiral modes. 

{\em Effective low-energy model.} The topological properties, including the nature of the edge states, can be better understood in terms of the following low-energy effective Hamiltonian (see 
Ref. \cite{sm}),
\begin{equation}
H^{\rm eff}_{\rm BdG}({\bf k})= d^0({\bf k}) \tau^0+ H^{\cal C}({\bf k}),    
\end{equation}
with  $d^0({\bf k})=-\alpha_R k_y  \xi_{\bf k}/E_{\bf k}$, $E_{\bf k}=\sqrt{\xi_{\bf k}^2+\Delta_0^2}$ and 
\begin{equation}\label{hcal}
    H^{{\cal C}}({\bf k}) = M({\bf k})\tau^x + \Delta_x k_x \tau^y,
\end{equation}
 being
$M({\bf k}) =  \left(V- E_{\bf k}\right)$, 
and $\Delta_x=\alpha_R \Delta_0/(2 E_{\bf k})$.
This representation makes explicit the chiral symmetry ${\cal C}\equiv \tau^z$ of the Hamiltonian $H^{{\cal C}}_{\bf k}$, according to which ${\cal C} H^{{\cal C}}({\bf k}) {\cal C}^{-1}= - H^{{\cal C}}({\bf k})$. Regarding $k_y$ as a parameter, we see that the topological phase
of the 1D limit ($k_y=0$) extends over a range of values $k_y$ satisfying $M(0,k_y)\geq 0$. We notice that Eq. (\ref{hcal}) is a Jackiw-Rebbi model \cite{Jackiw1976Jun,Goldstone1981Oct}, with an effective mass $M(x,k_y)$. This model has zero modes at the boundaries if a domain-wall profile is assumed in the $x$-boundary for the mass term. Furthermore, in Ref. \cite{sm} we explicitly calculate these edge modes and verify their Majorana nature. The extra term of $H^{{\rm eff}}_{\rm BdG}({\bf k})$ is proportional to $\tau^0$ and we can express in the boundary $d^0({\bf k})=v k_y$ with $v\simeq- \alpha_{\rm R} $.  Hence, the eigenstates of the effective Hamiltonian in the boundary are those of 
$H^{{\cal C}}({\bf k})$ with the energies shifted by $v k_y $ as stated in Eq. (\ref{h-edge}). 

\begin{figure}
\includegraphics[width=0.48\textwidth]{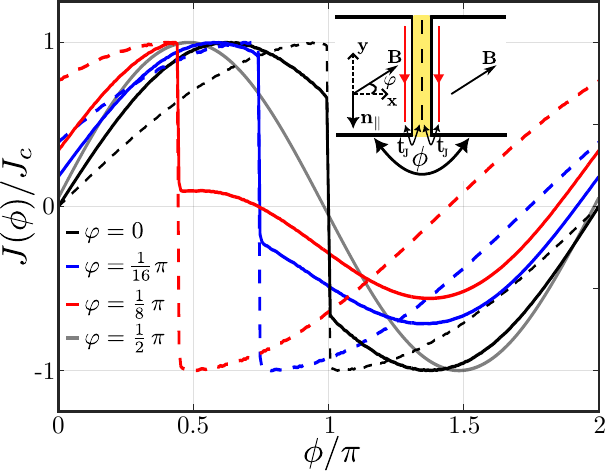}
\caption{\textbf{Josephson junction:} Current-phase relation (CPR) $J(\phi)$ relative to the critical current $J_c$ at zero temperature for different magnetic field orientations on the plane with polar angle $\varphi$ (see inset).  The  two superconductors are modeled by the lattice model with the same parameters as in Fig. \ref{fig:Wilson-loop}. The junction has 200 $k_y$ channels connected through 
 a row of sites with a hopping 
$t_{\rm J}=t/2$ (see Ref. \cite{sm}). The contribution to the CPR arising from the transverse mode with $k_y=0$ is shown with dashed lines. }
\label{fig:Josephson}
\end{figure}

{\em Josephson current.} The behavior of the the current-phase relation (CPR) $J(\phi)$ in the topological superconductor with antichiral Majorana modes is illustrated in Fig. \ref{fig:Josephson}. The  inset shows the
configurations for different angles $\varphi$ between the junction (${\bf n}_{||}$) and 
the magnetic field (${\bf n}_V$). Calculations were done considering 200 $k_y$-channels 
and details 
on the calculations are presented in Ref. \cite{sm}. For $\varphi$ within the topological phase (satisfying the condition of Eq. (\ref{angle})),
the supercurrent contribution associated to the zero mode $k_y=0$ is shown in dashed lines, where we can clearly identify the jump at $\phi=\pi$ when $\varphi=0$. We see that for $\varphi \neq 0$, the discontinuity in the supercurrent contribution corresponding to $k_y=0$ is shifted to $\phi\neq \pi$. 
Including all $k_y$-channels preserves the signature of the zero mode. 

The case $\varphi=0$ (see plots in black in Fig. \ref{fig:Josephson}) is akin the fully gapped 2D topological superconductors with many channels, the hybridization of the propagating Majorana modes can be described by an effective Dirac Hamiltonian with a $\phi$-dependent mass. The Andreev levels associated to the hybridization of these states with a tunneling amplitude ${\cal T}_{\rm J}\propto t_{\rm J}$ have energies $\varepsilon_{k_y}(\phi)=\pm \sqrt{(v k_y)^2 + {\cal T}_{\rm J}^2 \cos^2(\phi/2)} $ \cite{Ruiz2022Nov}. For $k_y=0$, because of the existence of Majorana zero modes, these two
states cross at $\phi=\pi$ and $J(\phi)$ has a jump at the value at
that value of the phase bias. This is, in turn, identical to the behavior of 1D systems with the magnetic field perpendicular to the SOC \cite{Lutchyn2010Aug,Oreg2010Oct,Kwon2004Feb,Pientka2013Feb}. The other 
 edge-channels with $k_y\neq 0$ have a smooth non-sinusoidal behavior. In contrast, the  $k_y$ channels corresponding to the bulk modes have a dispersion relation of the form $\varepsilon_{k_y}(\phi)=\pm E_{{\rm J}, k_y}\cos(\phi) $, being $E_{{\rm J}, k_y}$ a characteristic Josephson energy and  ontribute to the CPR as $\propto \sin(\phi)$. The total $J(\phi)$, results from the contribution of all the  Andreev states with negative energy. 
Therefore,  the overall shape and features of the CPR are determined by the relative spectral weight of the edge modes among the transverse channels, and by the degree of their hybridization compared to the transparency  for the bulk states. 

For other orientations of $\varphi\neq 0$ within the topological phase (see plots in red and blue in \ref{fig:Josephson}), the level crossing
of the component $k_y=0$, hence the jump 
in the CPR, occurs at different values of $\phi$ as in the case of 1D systems \cite{Spanslatt2018Aug,Aligia2020Dec,Cayao2024Feb}. For orientations $\varphi$ beyond the condition of Eq. (\ref{angle}) for the boundary of the topological phase,
the response is mainly dominated by a sinusoidal component for all the modes as in a non-topological superconductor (grey curve).

In all the cases, a response with $J(\phi=0)\neq 0$ is observed for $\varphi \neq 0$, typical of the anomalous Josephson effect. The amplitude of the critical current is also different for different signs of $\phi$, which  is the characteristic feature of  the Josephson diode effect \cite{Reynoso2008Sep,BaumgartnerPRL2021,BaumgartnerNature2022} observed in the non-topological phase.

{\em Conclusions.} We have shown the existence of a new topological phase in a 2D proximity-induced superconductor with Rashba spin-orbit coupling. This phase develops when a magnetic field is applied in-plane and is characterized by antichiral Majorana modes propagating along the edges perpendicular to the magnetic field. This phase takes place for a range of parameters (chemical potential and magnetic field) where the Bogoliubov-de Gennes spectrum has a single pair of Bogoliubov Fermi surfaces. We have shown that signatures of these modes can be identified in the behavior of the current-phase relation of a wide Josephson junction with the edges states oriented as sketched in Fig. \ref{fig:Josephson}. The topological phase persists for a range of angles close to this configuration. Other experimental signatures of these modes could be found in the heat transport \cite{Read2000Apr,Nomura2012Jan,Sumiyoshi2013Jan}, which is the counterpart to light propagation observed in photonic crystals with antichiral topological modes \cite{Zhou2020Dec,Chen2022Jun,Hu2024Nov}. Although these experiments are challenging in superconductors, they have been implemented to study heat propagation along topological edge states in spin liquids \cite{Kasahara2018Jul} and the quantum Hall state \cite{Granger2009Feb,Altimiras2012Jul}.

The experimental realization of this phase is possible in various material platforms. 
Natural candidates are Al/InAs heterostructures to confine a 2D electron gas with spin orbit and proximity-induced superconductivity. In these systems, high-transparency Josephson junctions have been demonstrated, exhibiting anomalous Josephson effect and diode effect, while supporting large in-plane magnetic fields ($\sim$1.5T) \cite{BaumgartnerPRL2021,BaumgartnerNature2022}. The typical electron density of the 2D gas is, albeit, not consistent with the range of chemical potentials for the topological phase. Nevertheless, a strategy can be  developed to control it by using two quantum wells: one of them buried deep in the heterostructure to be used as a back-gate and the other one very shallow to be proximitized by the thin superconducting film.  The gate tunability would allow to study the progressive effect of the emergence of pairs of Bogoliubov Fermi surfaces  until the conditions for the topological phase are reached. Another potential platform is a  two-dimensional magnetic topological insulator  like that studied in Ref. \cite{Mandal2024Jun}.

{\em Acknowledgements: } We thank Simon Feyrer, Nicola Paradiso, Luis Foa Torres, Jorge Facio,
Gerardo Ortiz and Shinsei Ryu for useful discussions. We acknowledge support from CONICET and FONCyT through PICT 2020-A-03661, Argentina. L.T. acknowledges the Georg Forster Fellowship from the Alexander von Humboldt Stiftung, Germany.

\bibliography{bibliography.bib}

\widetext
\clearpage

\begin{center}
\textbf{\large Supplementary Material:\\
Antichiral edge states and Bogoliubov Fermi surfaces in a two-dimensional proximity-induced superconductor} \\
Gabriel F. Rodríguez Ruiz$^1$,
Juan Herrera Mateos$^2$, Leandro Tosi$^3$,\\ Christoph Strunk$^4$, Carlos Balseiro$^1$, and Liliana Arrachea$^1$

\vspace{8pt}

$^1$\textit{Departamento de S\'olidos Cu\'anticos y Sistemas Desordenados, Centro At\'omico Bariloche, Instituto de Nanociencia y Nanotecnolog\'ia CONICET-CNEA and Instituto Balseiro (8400), San Carlos de Bariloche, Argentina.}

$^2$\textit{Instituto de Ciencias Físicas y Escuela de Ciencia y Tecnología, Universidad Nacional de San Martín, (1650) Buenos Aires, Argentina.}

$^3$\textit{Grupo de Circuitos Cuánticos Bariloche, Div. Dispositivos y Sensores, Centro Atómico Bariloche-CNEA, Instituto Balseiro and CONICET, (8400) San Carlos de Bariloche, Argentina.}

$^4$\textit{Institut f\"ur Experimentelle und Angewandte Physik, University of Regensburg, 93040 Regensburg, Germany.}
\end{center}

\setcounter{equation}{0}
\setcounter{figure}{0}
\setcounter{table}{0}
\setcounter{page}{1}
\makeatletter
\renewcommand{\theequation}{S\arabic{equation}}
\renewcommand{\thefigure}{S\arabic{figure}}

\section{Edge states in a disk}
The spatial probability distribution of the lowest-
positive-energy edge states in a disk is shown in Fig. 1. The finite size of the sample and the lattice discretization plays a role similar to disorder and generate interference
patterns. However, the edge states can still be identified.
\begin{figure}[h]
\includegraphics[width=0.48\textwidth]{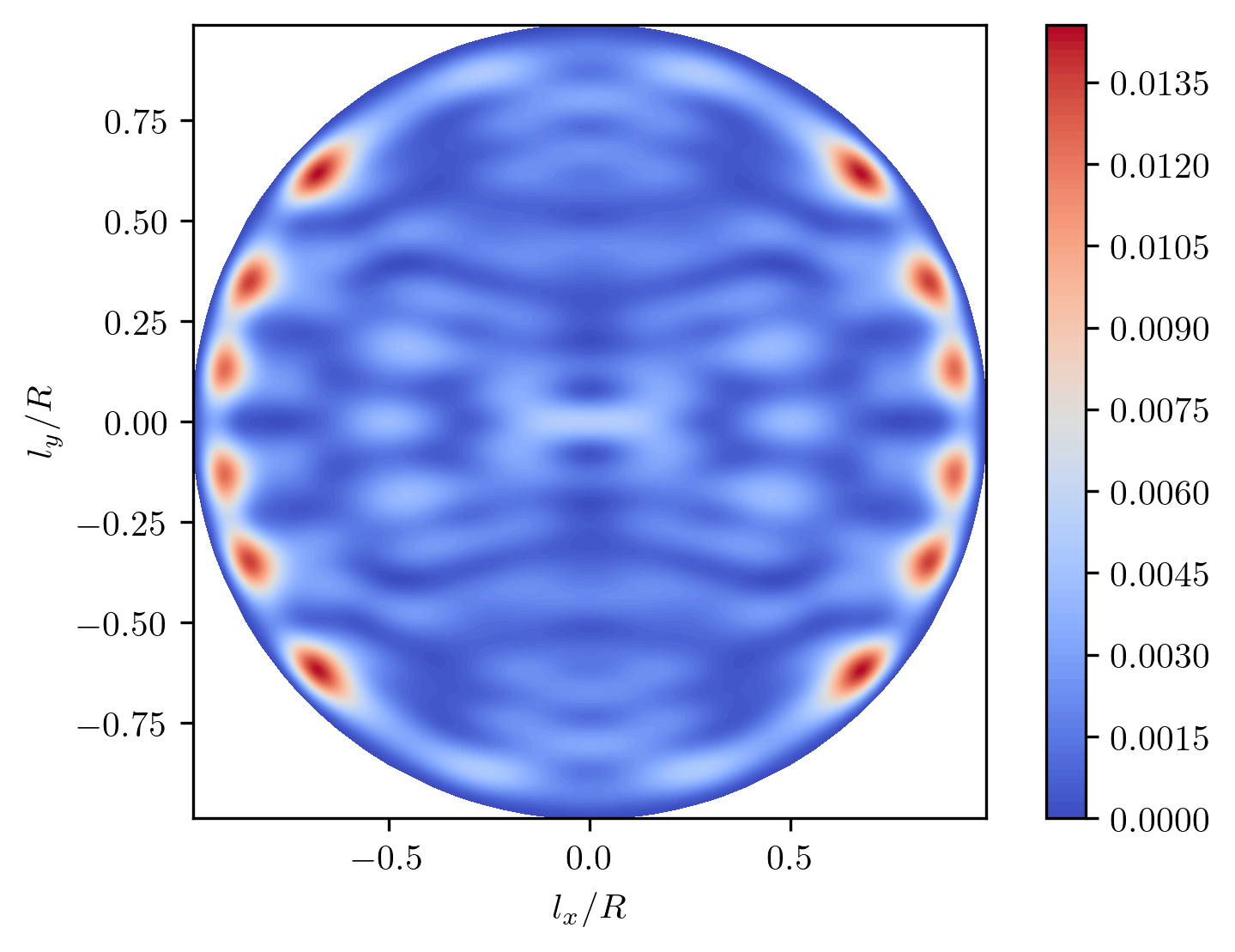}
\caption{\textbf{Disk symmetry:} Spatial probability distribution of the lowest positive energy eigenstate $E\approx0$ for an square lattice  with $N=100000$ sites embeded in a circular shape with radius $R=90$ sites, $V=(1.1\Delta_0, 0, 0)$, $E/t=1.7\times10^{-5}$. The rest of the parameters are the same as in the main text.}
\end{figure}

\section{Details of the derivation of the effective low-energy Hamiltonian}

We derive a low-energy Hamiltonian with a $2\times 2 $ matrix structure following two strategies.

\subsection{First order in the SOC}

We consider the  Hamiltonian with $\alpha_R=0$ and the quantization axis along the magnetic field. Defining the basis
${\bf c}_{\bf k}=\left(c_{\bf k,+},c_{{\bf k},-}\right)^T$, with $s=\pm$ denoting the spin parallel/antiparallel to the magnetic field, 
the Hamiltonian reads ${\cal H}_{0}({\bf k})=\overline{\cal H}({\bf k})+\overline{\cal H}(-{\bf k})$, being
\begin{eqnarray}\label{h1-2}
\overline{\cal H}({\bf k}) &=&\frac{1}{2} \left[ \left(\xi_{{\bf k}} - V\right)c^{\dagger}_{{\bf k},+}c_{{\bf k},+} +  \left(\xi_{{\bf k}} + V\right)c^{\dagger}_{-{\bf k},-} c_{-{\bf k},-}\right] 
- \frac{\Delta_0}{2}
\left(c^{\dagger}_{{\bf k},+} c^{\dagger}_{-{\bf k},-} + {\rm h.c.}\right).
\end{eqnarray}
This Hamiltonian is diagonalized by the transformation
\begin{equation*}\label{new-basis-2}
\left(\begin{array}c
\gamma_{{\bf k},+} \\
\gamma^{\dagger}_{-{\bf k},-}
\end{array}
\right)=
\left(
\begin{array}{cc}
u_{\bf k} & - v_{\bf k} \\
v_{\bf k}& u_{\bf k}
\end{array} \right)
\left(\begin{array}c
c_{{\bf k},+} \\
c^\dagger_{-{\bf k},-}
\end{array}\right),\;\;\;\;\;\;\;\;
\end{equation*}
\begin{equation*}
\left(\begin{array}c
c_{{\bf k},+} \\
c^{\dagger}_{-{\bf k},-}
\end{array}
\right)=
\left(
\begin{array}{cc}u_{\bf k} & v_{\bf k} \\
 -v_{\bf k}& u_{\bf k}
\end{array} \right)
\left(\begin{array}c
\gamma_{{\bf k},+} \\
\gamma^\dagger_{-{\bf k},-}
\end{array}\right),\;\;\;\;\;\;\;\;
\end{equation*}
with 
\begin{equation}
    u_{\bf k}^2=\frac{1}{2}\left(1+\frac{\xi_{\bf k}}{E_{\bf k}}\right),\;\;\;\;\;\;
    v_{\bf k}^2=\frac{1}{2}\left(1-\frac{\xi_{\bf k}}{E_{\bf k}}\right),
\end{equation}
and $E_{\bf k}=\sqrt{\xi_{\bf k}^2+\Delta_0^2}$.
Substituting in Eq. (\ref{h1-2}) we get (up to a constant)
\begin{equation}\label{hobog}
    {\cal H}_0({\bf k})= \sum_s \left(E_{\bf k}-s V \right) \left(\gamma^\dagger_{{\bf k},s}\gamma_{{\bf k},s} + \gamma^\dagger_{-{\bf k},s}\gamma_{-{\bf k},s}\right).
\end{equation}
We now focus on $V \gg \Delta$ and project the SOC term
\begin{equation} \label{soc-low}
{\cal H}_{\rm SOC} ({\bf k}) =-\frac{\alpha_{\rm R}}{2} \left(k_y +i k_x \right) \left(c^\dagger_{{\bf k}\uparrow}c_{{\bf k}\downarrow}  - c^\dagger_{-{\bf k}\uparrow}c_{{-\bf k}\downarrow}\right) +{\rm h.c.},
\end{equation}
on the lowest-energy band of the Hamiltonian Eq. (\ref{hobog}) (corresponding to $s=+$). 

Assuming the magnetic field along $x$ (${\bf  V}=V {\bf n}_x$), this implies
substituting in Eq. (\ref{soc-low})
\begin{equation}
    c_{{\bf k},\uparrow}= \frac{1}{\sqrt{2}}\left( c_{{\bf k},+}+ c_{{\bf k},-}\right)
    \rightarrow\frac{1}{\sqrt{2}}\left(u_{\bf k} \gamma_{{\bf k},+}-v_{\bf k}\gamma^\dagger_{-{\bf k},+}\right), \;\;\;\;\;c_{{\bf k},\downarrow}=\frac{1}{\sqrt{2}}\left( c_{{\bf k},+}- c_{{\bf k},-}\right)\rightarrow\frac{1}{\sqrt{2}}\left(u_{\bf k} \gamma_{{\bf k},+}+v_{\bf k}\gamma^\dagger_{-{\bf k},+}\right).
\end{equation}
After some algebra we get
\begin{equation}\label{soc-eff}
    {\cal H}_{\rm SOC}^{\rm eff} ({\bf k})=-\left(u_{\bf k}^2-v_{\bf k}^2\right)\alpha_{\rm R} k_y \left( \gamma^\dagger_{{\bf k},+} \gamma_{{\bf k},+}-
    \gamma^\dagger_{-{\bf k},+} \gamma_{-{\bf k},+}\right) - i \alpha_{\rm R} k_x  u_{\bf k} v_{\bf k}
    \left( \gamma^\dagger_{{\bf k},+} \gamma^\dagger_{-{\bf k},+}-
    \gamma_{-{\bf k},+} \gamma_{-{\bf k},+}\right).
\end{equation}
Combining Eqs. (\ref{hobog}) and (\ref{soc-eff}) and expressing this 
Hamiltonian in the basis $\Gamma({\bf k})=\left( \gamma_{{\bf k},+}, \gamma^\dagger_{-{\bf k},+}\right)^T$ as ${\cal H}_{\rm eff} ({\bf k})=\Gamma^\dagger({\bf k}) H_{\rm eff}^{\rm BdG}({\bf k}) \Gamma({\bf k}) $, we get the following BdG Hamiltonian
\begin{eqnarray}\label{heff0}
     H_{\rm BdG}^{\rm eff}({\bf k})&=& \left(E_{\bf k} -V\right)\tau^z + d^0({\bf k}) \tau^0 + \Delta_x k_x \tau^y,\nonumber \\
     d^0({\bf k}) &=&-\frac{\alpha_R \xi_{\bf k}}{E_{\bf k}}k_y, \;\;\;\;\;\;\; \Delta_x =\frac{\alpha_R \Delta_0}{2 E_{\bf k}}.
\end{eqnarray}
Here we see that there is an induced pairing with odd momentum dependence (p-wave type) in the direction of the magnetic field. 
Under the transformation $U^\dagger H U$, with $U=\left(\tau^0-i\tau^y\right)/\sqrt{2}$, this Hamiltonian can be expressed as
\begin{equation}\label{heff-1}
H^{\rm eff}_{\rm BdG}({\bf k})= H^{{\cal C}}({\bf k}) + d^0({\bf k}) \tau^0, 
    \end{equation}
    with 
\begin{equation}\label{hcal}
    H^{{\cal C}}({\bf k})= M({\bf k})\tau^x + \Delta_x k_x \tau^y, \;\;\;\;\;\;\;\; M({\bf k})= \left(V- E_{\bf k}\right).
\end{equation}

\subsection{Linear order in the pairing, linear order in the SOC}
Another route to derive the effective Hamiltonian  Eq. (\ref{heff0}) is to start by
 diagonalizing the normal part of the Hamiltonian
 including the kinetic, the SOC and the Zeeman fields. 
 
 Expressing the normal part of the Hamiltonian in
${\bf c}_{\bf k}=\left(c_{\bf k,\uparrow},c_{{\bf k},\downarrow}\right)^T$, it reads
 ${\cal H}_{\rm 2D}({\bf k})={\bf c}_{\bf k}^\dagger \; H_0({\bf k})\;{\bf c}_{\bf k}$, with
$H_0({\bf k})=   \xi_{\bf k}  +  H_{\rm SOC}({\bf k})+H_{\rm Z}$.
It is diagonalized by the transformation
\begin{equation*}\label{new-basis-1}
\left(\begin{array}c
c^{\dagger}_{{\bf k},+} \\
c^{\dagger}_{{\bf k},-}
\end{array}
\right)=\frac{1}{\sqrt{2}}
\left(
\begin{array}{cc}
1 & e^{i \theta_{\bf k}} \\
- e^{-i \theta_{\bf k}}& 1
\end{array} \right)
\left(\begin{array}c
c^\dagger_{{\bf k},\uparrow} \\
c^\dagger_{{\bf k},\downarrow}
\end{array}\right),
\end{equation*}
\begin{equation*}
\left(\begin{array}c
c^{\dagger}_{{\bf k},\uparrow} \\
c^{\dagger}_{{\bf k},\downarrow}
\end{array}
\right)=\frac{1}{\sqrt{2}}
\left(
\begin{array}{cc}
1 & -e^{i \theta_{\bf k}} \\
 e^{-i \theta_{\bf k}}& 1
\end{array} \right)
\left(\begin{array}c
c^\dagger_{{\bf k},+} \\
c^\dagger_{{\bf k},-}
\end{array}\right),
\end{equation*}
with $\theta_{\bf k} =\tan^{-1}\left(B_{\bf k}^y/B_{\bf k}^x\right)$ 
 and  ${\bf B}_{\bf k}= \left(B_{\bf k}^x, B_{\bf k}^y,0\right)$, with
 $B_{\bf k}^x=-V_x-\alpha_R k_y, \; B_{\bf k}^y= -V_y+\alpha_R k_x$. 
 The result reads
\begin{equation}\label{hcont2}
   {\cal H}_{\rm 2D}({\bf k})= \sum_{s=\pm } \xi_{{\bf k},s} c_{{\bf k},s}^\dagger  c_{{\bf k},s},
\end{equation}
with  $\xi_{{\bf k},\pm}=\xi_{\bf k} \pm |B_{\bf k}|$, and $|B_{\bf k}|=\sqrt{(B_{\bf k}^x)^2+(B_{\bf k}^y)^2}$.

The
pairing induced by proximity -- see the Hamiltonian  Eq. (2) of the main text --  expressed in the transformed basis reads
\begin{eqnarray}\label{pairing1}
{\cal H}_{\Delta}({\bf k})&=&  \Delta_{{\bf k},+} c^{\dagger}_{{\bf k},+}c^{\dagger}_{- {\bf k},+}+
\Delta_{{\bf k},-} c^{\dagger}_{{\bf k},-}c^{\dagger}_{- {\bf k},-}
+\Delta'_{{\bf k},}  c^{\dagger}_{{\bf k},+}c^{\dagger}_{-{\bf k},-} -\left(\Delta'_{{\bf k}}\right)^* c^{\dagger}_{{\bf k},-}c^{\dagger}_{-{\bf k},+}
+ {\rm h. c.} ,
\end{eqnarray}
with 
\begin{eqnarray}
\Delta_{{\bf k},+}&=&-\Delta_0(e^{-i  \theta_{-{\bf k}}}- e^{-i  \theta_{{\bf k}}})/4=-\Delta_0 \left(\cos \theta_{-{\bf k}}-\cos \theta_{{\bf k}} + i\sin \theta_{\bf k}-i\sin \theta_{-{\bf k}}\right)/4,\nonumber \\
\Delta_{{\bf k},-}&=&-\Delta_0(e^{i  \theta_{-{\bf k}}}- e^{i  \theta_{{\bf k}}})/4=-\Delta_0 \left(\cos \theta_{-{\bf k}}-\cos \theta_{{\bf k}} + i\sin \theta_{-{\bf k}}-i\sin \theta_{{\bf k}}\right)/4,\nonumber \\
\Delta'_{{\bf k}}&=&-\Delta_0(1+e^{-i \theta_{\bf k}} e^{i \theta_{-{\bf k}}})/4=-\Delta_0 \left[1+\cos (\theta_{{\bf k}}-\theta_{-{\bf k}})-i \sin(\theta_{{\bf k}}-\theta_{-{\bf k}})\right]/4.
\end{eqnarray}

Assuming the magnetic field along $x$ (${\bf  V}=V {\bf n}_x$) and $V>\alpha_R |{\bf k}|$, we can  approximate these expressions by keeping terms up to linear order in $k_x/V,\; k_y/V$ as follows,
$\tan(\theta_{\bf k}) \simeq -\alpha_R k_x/V$.
 The two pairing potentials read
\begin{equation}\label{eff-pair}
\Delta_{{\bf k},s}\simeq -i \frac{s \Delta_0 \alpha_R k_x}{2 V}, \;\;\;\;\;\;\;\;\;\;\;\;\;\;
\Delta'_{{\bf k}}\simeq  -\frac{\Delta_0}{2}\left(1+i \frac{ \alpha_R k_x}{V}\right).
\end{equation}


The relevant subspace for the topological phase corresponds to the lower band  $s=-$ of Eq. (\ref{hcont2}). We assume that the
Fermi energy is within this band. Hence, the 
dominant induced pairing is $\Delta_{{\bf k},-}$ defined in Eq. (\ref{pairing1}).
Therefore, the effective Hamiltonian reduces to
\begin{equation}
    \overline{\cal H}_{\rm eff}({\bf k})\simeq \left(\xi_{\bf k} - V-\alpha_R k_y \right)c_{{\bf k},-}^\dagger  c_{{\bf k},-}-
  \left( i \frac{\Delta_0 \alpha_R k_x}{2V} c^{\dagger}_{{\bf k},-}c^{\dagger}_{- {\bf k},-}+ {\rm h.c.}\right),
\end{equation}
where we have approximated $|B_{\bf k}|\simeq V+\alpha_R k_y$.
Introducing the Nambu basis ${\bf c}_{\bf k}=\left(c_{{\bf k},-},c^{\dagger}_{-{\bf k},-}\right)^T$, we can define
${\cal H}_{\rm eff}({\bf k})=\left[\overline{\cal H}_{\rm eff}({\bf k})+\overline{\cal H}_{\rm eff}({-\bf k})\right]/2$ and express it as follows ${\cal H}_{\rm eff}({\bf k})= {\bf c}^\dagger_{\bf k}{H}_{\rm BdG}^{\rm eff}({\bf k}){\bf c}_{\bf k}$, with
\begin{eqnarray}\label{heff-2}
    H_{\rm BdG}^{\rm eff}({\bf k}) &= & \left(\xi_{\bf k} - V \right)\tau^z  + \frac{\Delta_0 \alpha_R k_x}{2V}  \tau^y
    - \alpha_R k_y \tau^0.
\end{eqnarray}
 Recalling that the topological phase takes place for $\mu^2\leq V^2-\Delta_0^2$, we see that Eq. (\ref{heff-2}) agrees with
Eq. (\ref{heff0}) up to corrections $\propto \Delta_0^2$ within the range of parameters defining this phase.

\section{Calculation of the edge states in the effective continuum model}
We now show that the Hamiltonian Eq. (\ref{hcal}) regarded as $H_{\cal C}(x,k_y)$ has zero modes at the boundaries of the 
$x$ direction, within a certain range of $k_y$. To this 
end, we study the Jackiw-Rebbi Hamiltonian resulting from linearizing $H_{\cal C}({\bf k})$ in $k_x$ while keeping $k_y$ as a parameter,
\begin{equation}\label{dirac}
    H_{\rm JR}(k_y)=M(x,k_y)\tau^x -i\partial_x \Delta_x  \tau^y.
\end{equation}
We first notice that the topological phase exists for $M_0(k_y)=V-\sqrt{(\gamma k_y^2-\mu)^2+\Delta_0^2}\geq0$, being $\gamma=1/2m$.
This includes the special case $k_y=0$, where the model is equivalent to the 1D topological model for $V^2\geq \mu^2 +\Delta_0^2$, and 
extends over the range of $k_y$ satisfying 
\begin{equation}
    |k_y|\leq \frac{1}{\sqrt{\gamma}}\sqrt{\sqrt{V^2-\Delta_0^2}+\mu},
\end{equation}
which corresponds to the position of the cones in the spectrum.

To analyze the existence of a zero mode at the right boundary, we consider the topological phase in $x<0$ and assume a domain wall
at $x=0$ as follows: $M(x,k_y)=M_0(k_y) >0, \; x<0$,  and $M(x,k_y)=M_0(k_y) <0, \; x>0$.
We assume $\Delta_x>0$ and  look for a normalizable solution of
\begin{equation}
    \left[M_0(k_y)\tau^x -i\partial_x \Delta_x  \tau^y \right]\Psi_0(x,k_y)=0.
\end{equation}
The result is
\begin{equation}
    \Psi_{r}(x,k_y) =C \chi_r e^{\frac{M_0(k_y)}{\Delta_x} x},
\end{equation}
with $\chi_r=(1,0)^T$ and $C$ a normalization constant. Taking into account the transformation leading to Eq. (\ref{hcal}) we notice
that the operator associated to this spinor has the structure $\Gamma_r(k_y)=(\gamma_{k_y,+}-\gamma^\dagger_{-k_y,+})/\sqrt{2}$.
Multiplying by a phase factor, we can define the Majorana mode $\eta_{r, k_y}=i \Gamma_r(k_y)=\eta^{\dagger}_{r,-k_y}$.

Similarly, to analyze the existence of a zero mode at the left boundary, we consider the topological phase in $x>0$ and assume a domain wall
at $x=0$ as follows: $M(x,k_y) <0, \; x<0$,  and $M(x,k_y)=M_0(k_y)  >0, \; x>0$. The result is
\begin{equation}
    \Psi_l(x,k_y) =C \chi_l e^{-\frac{M_0(k_y)}{\Delta_x} x},
\end{equation}
with $\chi_l=(0,1)^T$. As in the case of the r zero mode, this spinor defines a Majorana fermion $\eta_{l, k_y}=(\gamma_{k_y,+}+\gamma^\dagger_{-k_y,+})/\sqrt{2}=
\eta^{\dagger}_{l,-k_y}$.

When we take into account the term $\propto \tau^0$ in Eqs. (\ref{heff-1}) and (\ref{heff-2}),  we
see that these edge modes have finite energy $v k_y$, with $v\simeq -\alpha_R$. 

\section{Details on the calculation of the Josephson current}

We calculate the Josephson current as a function of the flux $\Phi$ for a long junction with $N_y$ transverse channels, using the equilibrium Green's function formalism.

The system consists of two superconductors, $L$ and $R$, which are semi-infinite along the $x$- direction and modeled by a tight-binding Hamiltonian. The interface between the two superconductors is represented by a row of sites (see sketch of Fig.4 in the main text). Periodic boundary conditions are assumed along the $y$ direction, giving rise to $N_y$ transverse channels labeled by $k_y$.

The total Hamiltonian is given by
\begin{equation}\label{junction}
    {\cal H}=  {\cal H}_L+ {\cal H}_R+
    {\cal H}_{\rm J}(\phi),
\end{equation}
Where ${\cal H}_\alpha$ with $\alpha=L,R$ are the Hamiltonians for the two superconductors. Introducing  the Nambu basis ${\bf c}_{\alpha, l_x,k_y} =\left( c_{\alpha, l_x,k_y,\uparrow}, c_{\alpha, l_x,k_y,\downarrow},c^{\dagger}_{\alpha, l_x,-k_y,\downarrow}, -c^{\dagger}_{\alpha, l_x,-k_y,\uparrow} \right)^T$
they can be written as follows
\begin{eqnarray}
{\cal H}_\alpha&=&  \frac{1}{2} \sum_{l_x=1}^{\infty} \left[ {\bf c}^{\dagger}_{\alpha, l_x,k_y} \left[ \tau^z(\xi_{k_y}-2\lambda\,\sin k_y\,\sigma^x)-V\mathbf{n_V}\cdot\mathbf{\sigma}+\Delta_0\tau^x \right]{\bf c}_{\alpha, l_x,k_y} \right.\nonumber \\
& & \left.
+{\bf c}^{\dagger}_{\alpha,l_x,k_y} \left[\tau^z(-t-is_\alpha \lambda\sigma^y)
\right]{\bf c}_{\alpha, l_x+1,k_y}+\text{h.c.}
\right],
\end{eqnarray}
where $j_x$ counts rows of sites along $x$, with $l_x=1$ being the row closest to the junction
and $s_R=-s_L=1$. The other term of Eq. (\ref{junction}) describes the junction. We assume it has an interface represented by a line of $N_y$ normal sites to which the line of sites at the 
edge of each superconductor is connected through a hopping term $t_{\rm J}$. This term reads
\begin{equation}
    {\cal H}_{\rm J}(\phi) =  \sum_{k_y,\sigma } \left[t_{\rm J}\left(e^{i\phi/4} c^\dagger_{L,1, k_y \sigma} d_{k_y,\sigma} + e^{i\phi/4} d^\dagger_{k_y\sigma} c_{R,1,k_y \sigma} +{\rm h.c.}\right)+\xi_{d,k_y} d^\dagger_{k_y,\sigma}d_{k_y,\sigma}\right]
\end{equation}
with $\phi=2\pi \Phi/\Phi_0$, being $\Phi_0$ the flux quantum and $\xi_{d,k_y}=-\mu+4t-2t\cos k_y$.





We introduce the Nambu basis for the interface ${\bf d}_{k_y} =\left( d_{k_y,\uparrow}, d_{k_y,\downarrow},d^{\dagger}_{-k_y,\downarrow}, -d^{\dagger}_{-k_y,\uparrow} \right)^T$ 
and express the hopping matrix in the Bogoliubov de Gennes representation as follows
\begin{equation}
\hat{\tau}(\phi) = t_{\rm J} \left[e^{i\phi/4}\left(\tau^z+\tau^0\right)  +e^{-i\phi/4}\left(\tau^z-\tau^0\right) \right]\frac{\sigma^0}{2}.
\end{equation}

The Josephson current is then expressed as
\begin{equation}
J(\phi)=\frac{e}{\hbar}\sum_{k_y}\mbox{Re} \left\{ \mbox{Tr} \left[\tau^z  \sigma^0 \; \hat{\tau}(\phi){\bf G}^<_{d,L}(k_y;t,t) \right]\right\}=\frac{e }{h}\sum_{k_y}\mbox{Re} \left\{\int  d\varepsilon \mbox{Tr} \left[\tau^z  \sigma^0 \; \hat{\tau}(\phi){\bf G}^<_{d,L}(k_y,\varepsilon) \right]\right\},    
\end{equation}

where we have introduced the lesser Green's function
\begin{equation}
{\bf G}^<_{d,L}(t,t^{\prime}) =-i\langle {\bf c}_{L,1,k_y}^{\dagger}(t^{\prime}) {\bf d}_{k_y}(t) \rangle,
\end{equation}
and its Fourier transform $t-t^{\prime} \rightarrow \varepsilon$. Using Langreth rules, we obtain
\begin{equation}
{\bf G}^<_{d,L}(k_y,\varepsilon)={\bf G}^<_{d,d}(k_y,\varepsilon)\hat{\tau}^\dagger(\phi){\bf g}^{a}_{L}(k_y,\varepsilon)+{\bf G}^r_{d,d}(k_y,\varepsilon)\hat{\tau}^\dagger(\phi){\bf g}^<_{L}(k_y,\varepsilon),    
\end{equation}
where we have introduced the retarded Green’s functions
\begin{equation}
{\bf G}^r_{d,d}(k_y,\varepsilon)=[{\bf g}^r_{d,d}(k_y,\varepsilon)^{-1}-{\bf \Sigma}^r_{\rm L}(k_y,\varepsilon)-{\bf \Sigma}^r_{\rm R}(k_y,\varepsilon)]^{-1},
\end{equation}
and 
\begin{equation}
{\bf g}^r_{d,d}(k_y,\varepsilon)=[\varepsilon\tau^0-\xi_{k_y}\tau^z]^{-1},
\end{equation}
The self-energies are defined as
\begin{align}
\boldsymbol{\Sigma}^r_{L}(k_y,\varepsilon) & = \hat{\tau}(\phi)^\dagger{\bf g}^{r}_{L}(k_y,\varepsilon) \hat{\tau}(\phi),
\\
\boldsymbol{\Sigma}^r_{\rm R}(k_y,\varepsilon) & = \hat{\tau}(\phi)\,{\bf g}^{r}_{R}(k_y,\varepsilon) \hat{\tau}^\dagger(\phi),
\end{align}
being ${\bf g}^{r}_{\alpha}$ ($\alpha=L,R$)  the surface Green's function of the superconductor at the site adjacent to the interface. This is computed using the same recursive algorithm as in Ref. \cite{Gruneiro2023Jul,Mateos2024Aug}. The advanced Green’s functions are obtained via ${\bf G}_{ij}^a(k_y,\varepsilon)= \left[{\bf G}^r_{ji}(k_y,\varepsilon)\right]^{\dagger}$.

In equilibrium, the lesser Green's functions are given by:
\begin{align}
{\bf G}^<_{d,d}(k_y,\varepsilon)&=f(\varepsilon)\left[{\bf G}^a_{d,d}(k_y,\varepsilon)-{\bf G}^r_{d,d}(k_y,\varepsilon) \right],\\
{\bf g}^<_{L}(k_y,\varepsilon)&=f(\varepsilon)\left[{\bf g}^a_{L}(k_y,\varepsilon)-{\bf g}^r_{L}(k_y,\varepsilon) \right],    
\end{align}
so the Josephson current becomes
\begin{equation}
J(\phi)=\frac{e}{h}\sum_{k_y} \left\{ \int d\varepsilon f(\varepsilon)F_{k_y}(\varepsilon)\right\},
\end{equation}
with
\begin{equation}
F_{k_y}(\varepsilon)=\mbox{Re} \left\{ \mbox{Tr}\left[\tau^z\hat{\tau}(\phi)\left({\bf G}^a_{d,d}(k_y,\varepsilon)\hat{\tau}^\dagger(\phi){\bf g}^{a}_{L}(k_y,\varepsilon)+{\bf G}^r_{d,d}(k_y,\varepsilon)\hat{\tau}^\dagger(\phi){\bf g}^r_{L}(k_y,\varepsilon)
\right)\right] \right\}.
\end{equation}


All calculations are carried out at zero temperature.

\end{document}